\documentclass[
 aip,
amsmath,amssymb,
preprint,
]{revtex4-1}

\usepackage{graphicx}
\usepackage{dcolumn}
\usepackage{bm}

\usepackage[utf8]{inputenc}
\usepackage[T1]{fontenc}
\usepackage{mathptmx}
\usepackage{etoolbox}
 \usepackage{color}
 \usepackage{hyperref}

\makeatletter
\def\@email#1#2{%
 \endgroup
 \patchcmd{\titleblock@produce}
  {\frontmatter@RRAPformat}
  {\frontmatter@RRAPformat{\produce@RRAP{*#1\href{mailto:#2}{#2}}}\frontmatter@RRAPformat}
  {}{}
}%
\makeatother
\begin{document}
\preprint{}
\title{Charged drop impinging on particles dispersed over a  metallic plate:  A method of self-cleaning}

\author{D. Biswal}
 \affiliation{Department of Mechanical Engineering, Indian Institute of Technology Delhi, New Delhi - 110016, India.}

 \author{S. K. Saroj}%
\affiliation{ 
Gulliver and PMMH Laboratories, CNRS, ESPCI Paris, Université PSL, 75005, Paris, France
}

\author{B. Ray}
\email{bray@mech.iitd.ac.in}
\affiliation{Department of Mechanical Engineering, Indian Institute of Technology Delhi, New Delhi - 110016, India.}

\author{Debabrata Dasgupta}
\affiliation{Department of Mechanical Engineering, Indian Institute of Technology Delhi, New Delhi - 110016, India.}

\author{R. M. Thaokar}
\affiliation{Department of Chemical Engineering, Indian Institute of Technology Bombay, Mumbai - 400076, India.}

\author{Y.S. Mayya}
\affiliation{Department of Chemical Engineering, Indian Institute of Technology Bombay, Mumbai - 400076, India.}

\begin{abstract}
\section*{Abstract}
 An electric field applied to a droplet impinging on a hydrophobic surface has an extensive variety of applications, including ant-icing, heat transfer enhancement, self-cleaning, droplet manipulation, and electrostatic spraying. The present study demonstrates an effective method of particle removal using a charged droplet. This method employs a pin-plate electrode setup to investigate the dynamics of a charged droplet impact on the surface covered with particles. The particles of different properties such as wettability, electrical conductivity, etc. have been used. Silane-coated glass beads, carbon black, and glass beads are dispersed over the ground copper electrode. The applied potential is also varied from 2 kV to 4 kV. A high-speed imaging is employed to visualize the drop motion, dynamic behavior, and self-cleaning phenomenon. The experimental results indicate that drop generation and impact occur at applied potentials of 2.5, 3, and 3.5 kV, in contrast, at 2 kV, there is no droplet pinch-off. At 4 kV, electric breakdown and bridging of the droplet between the capillary and ground electrode are observed. The drop impact on the silane-coated glass bead leads to their attachment due to the adhesiveness of the particles and the droplet. The silane-coated particles are removed from the droplet surface due to the deformation of the drop and the electric repulsive force. In the case of carbon black and glass beads, the particles are captured by the droplet due to the electrostatic force of attraction. Higher electric potentials lead to an increased spreading diameter of the droplet. The higher electric field enhances the contact area between the droplet and the particles, thereby removing more particles.
\end{abstract}

\maketitle
\section{\label{sec:1:introduction}Introduction}
The impact of charged droplet under electric field has a significant scientific interest as it has a wide variety of applications: condensation heat transfer enhancement \cite{leach2006dropwise}, anti-icing surface \cite{wang2015superhydrophobic, farhadi2011anti}, electrostatic spraying, ant-frosting \cite{jing2013frosting}, inkjet printing \cite{link1997fluidized} and self-cleaning \cite{furstner2005wetting, wisdom2013self}.  The impact and dynamic behavior of droplets have been widely carried out using superhydrophobic surfaces. Superhydrophobic surfaces are generally made using chemicals, surface coating, etc \cite{wang2007impact}. When the drop falls on a superhydrophobic surface, a droplet impact can result in deposition, splashing, breakup, partial rebound, or complete rebound phenomena depending on the impact parameters and surface properties \cite{yarin2006drop}. The drop impact depends on: (1): fluid and physical properties: density, surface tension, viscosity, velocity, charge, and droplet diameter, (2) surface properties: temperature, wettability, inclination, and surface roughness \cite{marengo2011drop}. The impact and dynamic behavior of the droplet have been explained through some critical parameters: spreading factor \cite{xu2017maximum, arogeti2019drop}, non-dimensional numbers, and contact time \cite{tsai2009drop, malgarinos2014vof}. The maximum spreading factor is a critical parameter to understand the impact of a droplet on solid surfaces. It is defined as the ratio of the maximum spreading diameter to the initial droplet diameter before impact \cite{xu2017maximum, arogeti2019drop}.\\
The use of an external electric field to study the droplet impact behaviors has gained significant attention. The electric field is typically applied through three different charging methods: electrowetting-on-dielectric (EWOD), impacting within the electric field, or pre-charging the droplet before impact \cite{xu2021impact}. A numerous experimental and numerical investigations on drop impact and dynamics have been done to understand the underlined physics, which are delineated below.\\\\
\citeauthor{malouin2010directed} observed that the drop rebounding phenomenon on textured surfaces can be significantly affected by the uniformity of the surface roughness. \citeauthor{rioboo2008drop} observed the Cassie-Baxter/Wenzel transition follows the function of the drop size, as well as drop deposition, rebound, sticking, or fragmentation. \citeauthor{han2022droplet} reviewed the fundamentals, regulations, and applications of millimeter-sized droplets bouncing on solid surfaces and same/miscible liquids. \citeauthor{yun2014ellipsoidal} conducted both experimental and numerical investigations, concentrating on the impact of ellipsoidal drops on a non-wetting surface. Their research delved into the influence of aspect ratio and Weber number on drop dynamics. The experimental findings revealed that a higher aspect ratio resulted in more pronounced impact behavior, impeding droplet rebound. Additionally, they noted that droplet recoiling and rebounding were influenced by both the Weber number and aspect ratio. \citeauthor{werner2007droplet} studied the impact and spreading of droplets on the surface of anhydrous milkfat. Their findings revealed that a greater impact velocity and lower viscosity of droplets resulted in the maximum spread diameter. Additionally, the introduction of a surfactant did not alter the spreading diameter but effectively prevented droplet recoiling compared to a liquid without the surfactant. \citeauthor{zhang1997dynamic} investigated how the dynamics of a droplet impacting a flat glass surface are influenced by dynamic surface tension. They found that dynamic surface tension plays a role in reducing surface energy, leading to a maximum spreading diameter. Additionally, they noted that an uneven distribution of dynamic surface tension generates Marangoni stresses, impeding the spreading of the droplet. \citeauthor{han2022droplet} reviewed the fundamentals, regulations, and applications of millimeter-sized droplets bouncing on solid surfaces and same/miscible liquids. \citeauthor{phenomena2002contact} evaluated a correlation for contact time by balancing inertial and capillary terms for a wide range of radii (0.1– 4 mm) as follows:
\begin{equation}
    \tau \sim (\frac{\rho R^3}{\sigma})^{1/2}
\end{equation}
\citeauthor{li2017electrohydrodynamic} investigated the electrohydrodynamic characteristics of water droplets that propel themselves and undergo breakup under the influence of a tangential electric field, surface microstructure, and chemical composition. Additionally, they observed how a charged sessile drop, driven by the tangential electric field, helps in the self-cleaning of particle contaminants on a superhydrophobic surface. \citeauthor{traipattanakul2017study} investigated the motion of a charged droplet, the threshold electric field, and the charge required by a droplet to remove a macro-sized droplet from the superhydrophobic surface using uniform electrode configurations. \citeauthor{miljkovic2014jumping} demonstrated the extraction of electrostatic energy by employing bouncing droplets between two parallel surfaces with top: superhydrophobic and bottom: hydrophilic. \citeauthor{mukherjee2021electrostatic} demonstrated the jumping and braking of micrometric frost growing on the chilled surface using polarizable liquid under an electric field. \citeauthor{mhatre2013drop} investigated the movement and shape change of a conductive droplet within an imperfect or permeable dielectric liquid subjected to a non-uniform electric field. They elucidated the dielectrophoretic motion of the conductive droplet in both perfect and leaky dielectric mediums. Their findings revealed that the non-uniform electric field is effective in inducing droplet deformation and breakup, surpassing the effects of a uniform electric field. \citeauthor{zhang1996dynamics} examined how an electric field influences the dynamics of drop formation in the dripping mode from a vertical metal capillary containing viscous liquid. Their investigation revealed that as the electric field strength increased, the drop volume decreased, while the length of the jetting thread increased. \citeauthor{ferrera2013dynamical} investigated the electrodynamic characteristics of low-conductivity droplets under electrified conditions through both experimental and numerical approaches. The results revealed a consistent agreement between the experimental and numerical studies, particularly in the transitional regime of drop-jet behavior.  \citeauthor{traipattanakul2019electrostatic} examined the jumping height, the charge, and the force exerted on the droplets through experimentation and mathematical modeling. Their findings revealed that the application of an electric field could triple the height of the droplets compared to those that were uncharged. \citeauthor{tian2022does} conducted both experimental and numerical analyses to explore how the electrostatic force influences the ejection and rebounding of droplets on a superhydrophobic surface. The research encompassed the examination of four ejection regimes and three rebounding regimes. Their findings indicated that the mode of ejection is determined by the force exerted on the droplet apex. \citeauthor{miljkovic2013electrostatic} investigated how jumping droplets exhibit a repulsive force due to their net positive charge. Additionally, they measured the overall charge of the droplets and investigated the mechanisms behind both the repelling and jumping behavior of the droplets.  \citeauthor{deng2023behaviors} conducted research on how temperature and different electric fields influence the impact of drops on a superhydrophobic surface. The focus was on analyzing drop contact time, maximum spreading factor, rebound, and nucleation rate in relation to varying surface temperatures and electric fields. The findings revealed that the dynamic behavior of water droplets during retraction and rebound phases is influenced by both surface temperature and electric field. \citeauthor{takeda2002control} studied that the motion of the droplet can be controlled by using the static electric field. They concluded that the reduction of the applied electric field and droplet falling height can hinder the droplet motion. \\ \\
The existing literature reveals a substantial amount of experimental work and numerical analysis on the dynamic behavior of drop impact. The prevailing trend in existing literature primarily focuses on drop impact investigations involving the making of superhydrophobic surfaces, the use of chemicals, and uniform electric fields. Notably, there is a conspicuous absence of discussions in the literature concerning the dynamic of drop impact and bouncing without resorting to the creation of superhydrophobic surfaces. It is imperative to delve into the fundamental physics underlying the dynamics of drop impact with self-cleaning of micron particles from non-hydrophobic surfaces using the non-uniform electric field.\\ \\
Motivated by the self-cleaning application using the drop impact and bouncing method, this study delves into the dynamic aspects of drop impact including spreading, recoiling, and bouncing phenomena on the metallic ground electrode. This dynamic aspect drop impact is studied under a non-uniform electric field at different applied potentials. The ground metallic electrode is covered with micron-sized particles with different electrical and chemical properties. This study also emphasizes how the particles spread on the metallic ground electrode, can be removed using the charged drop impact and bouncing method.

\section{\label{sec:2:experimental setup}Experimental details and methodology}
\subsection{\label{sec:2:experimental setup}Experimental setup}
\begin{figure*}
    \centering
    \includegraphics[width=0.8\linewidth]{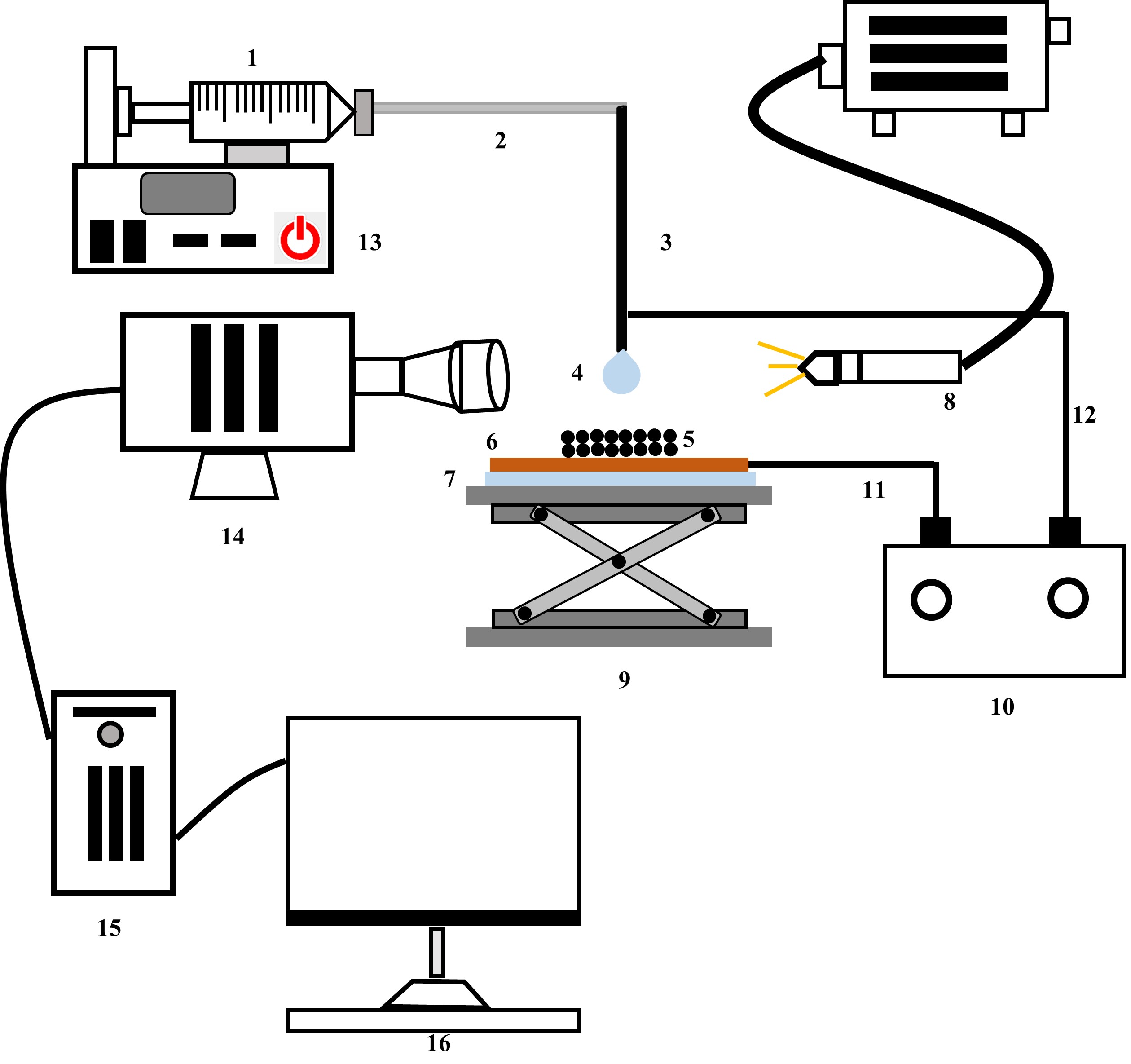}
    \caption{Layout of the experimental setup: (1) 10 ml plastic syringe, (2) 3 mm Teflon flowing pipe, (3) metallic needle, (4) pendant droplet, (5) particles, (6) copper plate, (7) 5 mm thickness acrylic sheet, (8) LED light, (9) laboratory jack, (10) DC power source, (11) ground electrode of DC source, (12) live electrode of DC source, (13) syringe pump, (14) high-speed camera, (15) CPU of the computer, (16) computer}
    \label{fig:1}
\end{figure*}

Figure \ref{fig:1} shows the schematic of the experimental setup to investigate the behavior of the charged droplet in a non-uniform electric field while cleaning the particles from the ground electrode. The setup consists of a metallic capillary for suspending the pendant charged droplet. Length, outer diameter (OD), and inner diameter (ID) of the capillary are 38.1 mm, 0.72 mm, and 0.41 mm, respectively. The capillary is used as one of the electrodes and is vertically attached to an insulated holder. The insulated holder is attached to the metallic burette stand. A copper plate (1 mm thick, 40 mm wide, and 80 mm long) used as the ground electrode is attached to a 10 mm thick acrylic sheet and kept on a laboratory jack. The distance between the ground electrode and the tip of the needle can be adjusted by moving the jack up or down in a vertical direction. A high voltage DC power source (10 kV, 10 mA, dual polarity, Ionics Pvt. ltd) is used to charge the droplet. One electrode of the DC source is connected to the capillary, while the other is grounded to the copper plate. The syringe pump (Cole–Parmer) with a 10 ml plastic syringe is employed to generate the pendant drop at the tip of the capillary. The Milli – Q water (18.2 M$\Omega$cm) used as a working liquid is pumped to the capillary with the help of a flow pipe (3 mm ID).

The hollow glass beads (Sigmaaldrich), silane-coated glass beads (Cospheric), and super C65 carbon black (C-NERGY$^{TM}$) particles are used in the current study. The physical properties of the particles are provided in Table \ref{tab:table1}.
\begin{table*}
\caption{\label{tab:table1} Properties of particles: P1: hollow glass beads; P2: carbon black; P3: silane coated glass beads; $\rho_p$: density; $\sigma_p$: conductivity; $\epsilon_{rp}$: relative permittivity}
\begin{ruledtabular}
\begin{tabular}{cccddd}
Particles & $D_m$ ($\mu$ m) & $\rho$ (g/cc) & \sigma (S/m) & wettability & \epsilon_{rp} \\
\hline
P1 & 9-13 & 1.1 - 1.3 & 10^{-11} & Good & 6\\
P2 & 0.1-0.5 & 1.6 & 2\times 10^2 & Poor & 15\\
P3 & 5-9 & 2.5 & 10^{-16} & Poor & 32\\
\end{tabular}
\end{ruledtabular}
\end{table*}

\subsection{\label{sec:2:experimental setup}Methodology}
In this experiment, a certain amount of particles are distributed onto a copper plate. The distance between the tip of the capillary and the copper plate is kept equal to 6 mm. A syringe pump generates the pendant droplet at the tip of the capillary with a controlled flow rate of 10 $\mu$l/min. This pendant droplet is charged by keeping the needle as the positive electrode and the copper plate is grounded. The droplet pinched off from the capillary is observed for the applied potential in the range of 2.5 kV to 4kV. To visualize and record the dynamics of the charged droplet along the particle cleaning on the surface, a high-speed camera (Phantom V12) equipped with a microscope (Stereo microscope, Leica) is employed.  An LED light source illuminates the interaction zone and its surroundings. The resolution and camera speed are 1280 $\times$ 800 and 5000 frames per second, respectively. 
\begin{figure*}
    \centering
    \includegraphics[width=0.9\linewidth]{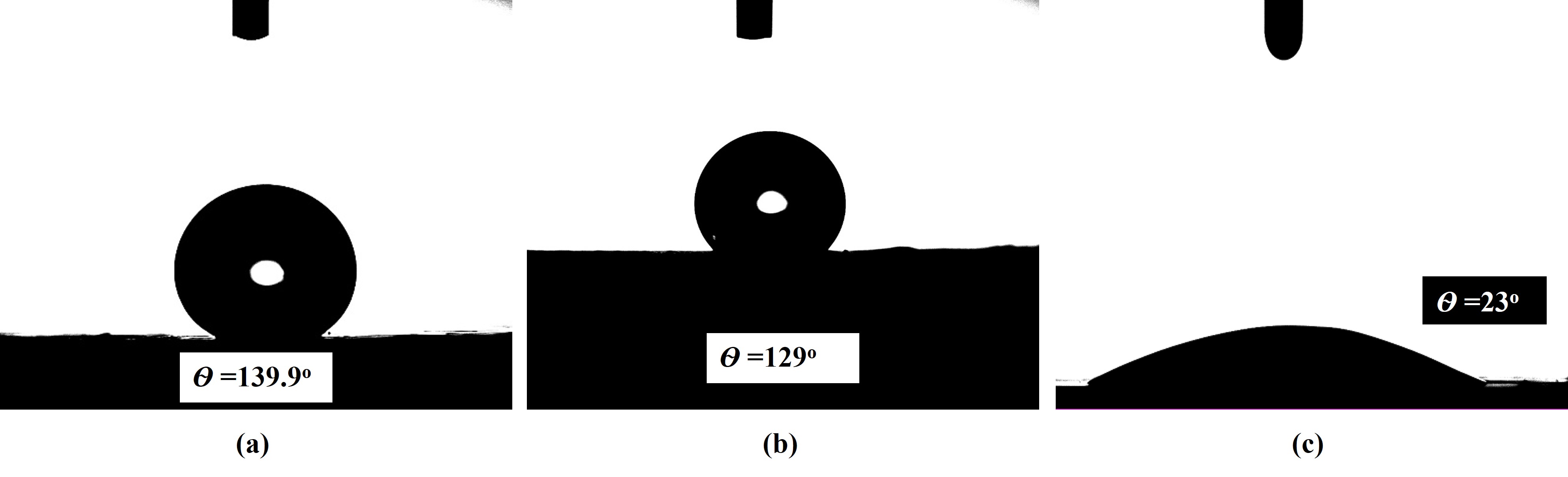}
    \caption{Contact angle of the droplet placed on the particle surface: (a) carbon black, (b) silane-coated glass beads, and (c)} glass beads.
    \label{fig:2}
\end{figure*}
\subsection{\label{sec:2.1}Measurement of wettability effect of three different particles}
As the dynamics of the drop impact depend on the surface properties of the substrate, the wettability of hollow glass beads, carbon black, and silane-coated glass beads is measured in terms of contact angle. To measure the wettability of the particle surface, a 4 mm-sized pellet is formed using a palletizer, applying manual pressure to ensure its integrity without any cracks/pores. These pellets are affixed to a goniometer (DIGIDROP, GBX) for Young’s contact angle measurement. The contact angle is determined by vertically positioning a sessile droplet of deionized (DI) water with a fixed volume of 10 $\mu$l onto the pellet's surface. Using a goniometer and monochromatic camera, images of the sessile droplet resting on the pellet are captured. Three sets of measurements are conducted at three different locations on the pellet, and the average contact angle is determined. The results indicate that carbon black particles exhibit a very low wettability effect with a contact angle of 140$^\circ$, surpassing silane-coated glass beads with a contact angle of 130$^\circ$ and hollow glass beads with a contact angle of 23$^\circ$ as shown in Figure \ref{fig:2}. 

\section{\label{sec:3:Results and discussions}Results and discussions}
This section includes the experimental observation of the dynamics of the charged droplet impacting on the particles dispersed on the ground electrode. The experiments have been conducted as function of the applied potential and type of particles. The results of impacting dynamics of droplet in the absence of applied potential are also included to demonstrate the contrast with the dynamics of droplet in the presence of applied potential.


\subsection{Dynamics of drop impact on the particle surface without effect of electric field}
\label{absence of electric field}
\begin{figure*}
    \centering
    \includegraphics[width=0.8\linewidth]{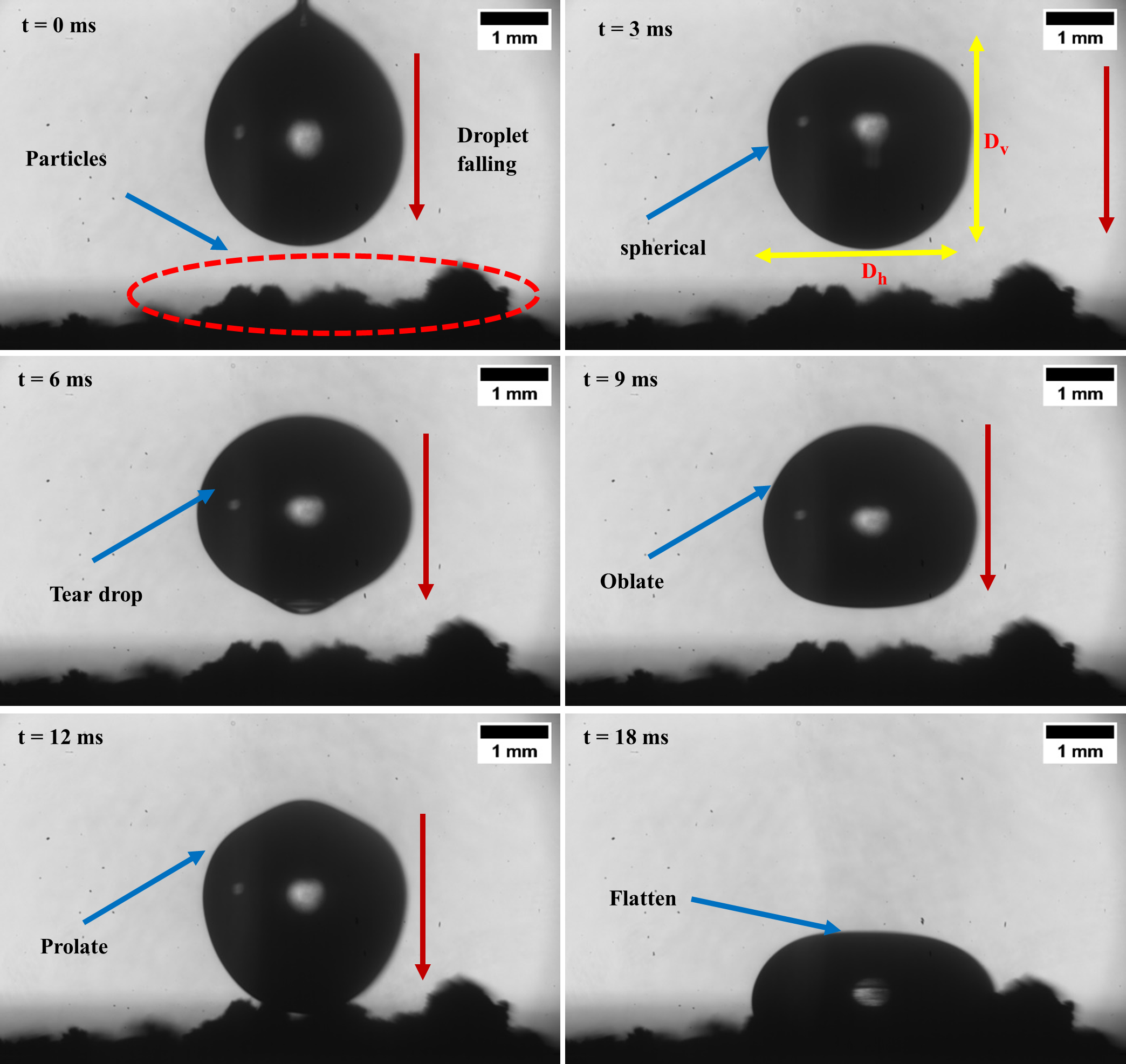}
    \caption{Time sequential images showing the dynamics of drop on the substrate covered with silane-coated glass beads in the absence of electric field. The corresponding video is available as video 1 in SI. {\color{blue}\href{https://drive.google.com/file/d/1RACS8Qrs1ZNd-S521CcmDyNPQOwK2Xod/view?usp=drive_link}{video1}}}
    \label{fig:without_charge}
\end{figure*}

In this case, the droplet impacts the particle's surface in the absence of the electric field. Here the silane-coated glass beads are dispersed on the copper ground electrode. The flow rate of the syringe pump is maintained at 50 $\mu$l/min for generating a pendant droplet. At such a flow rate of water, the droplet generates and grows at the tip of the capillary (see Figure \ref{fig:without_charge} at time (t) = 0 ms). The syringe pump is kept on running and the water continues to flow until the droplet separates from the capillary. When the inertia of flow and size of the droplet exceeds a certain range, inertia and gravity overcome the surface tension force. The droplet pinches off from the capillary at t = 3 ms and starts moving towards the surface of the particles (see Figure \ref{fig:without_charge} at t = 3 ms). During the downward motion, the droplet's size is measured to be approximately 2.8 mm. The droplet achieves a maximum velocity of 0.35 m/s before touching the surface. Observations show that upon impact at t= 12 ms, some particles adhere to the droplet's surface, suggesting a partial hydrophobic nature of the droplet. Despite the droplet gaining some surface energy, this energy is insufficient to cause the droplet to bounce back from the surface. Instead of the droplet bouncing back from the surface, it recoils on the particle surface. After a certain time on the particle surface, the droplet maximizes its contact area and settles on the particle surface (see Figure \ref{fig:without_charge} at t = 18 ms). This finding indicates that although some particles are captured on the surface of the uncharged droplet upon impact, the droplet remains on the target surface rather than moving away with the particles from that area. It indicates that a moving uncharged droplet is ineffective in cleaning stationary dust from a specific surface. 
\subsection{Effect of electric field on drop impact on the particle surface}
As illustrated in the earlier section \ref{absence of electric field}, the droplet collects some of the particles during striking on the surface, but it does not effectively clean the target surface. Therefore, the electric field is applied to charge the droplet. This cleans the target surface by changing the impacting dynamics of the droplet depending on the strength of the electric field and the type of particles. The experimental observation as a function of the electric field for different types of particles is described in this section.

\subsubsection{\label{sec:3.2}Dynamics of droplet impacting on the highly conducting and hydrophobic particles (carbon black)}
\begin{figure*}
    \centering
    \includegraphics[width= 1\linewidth]{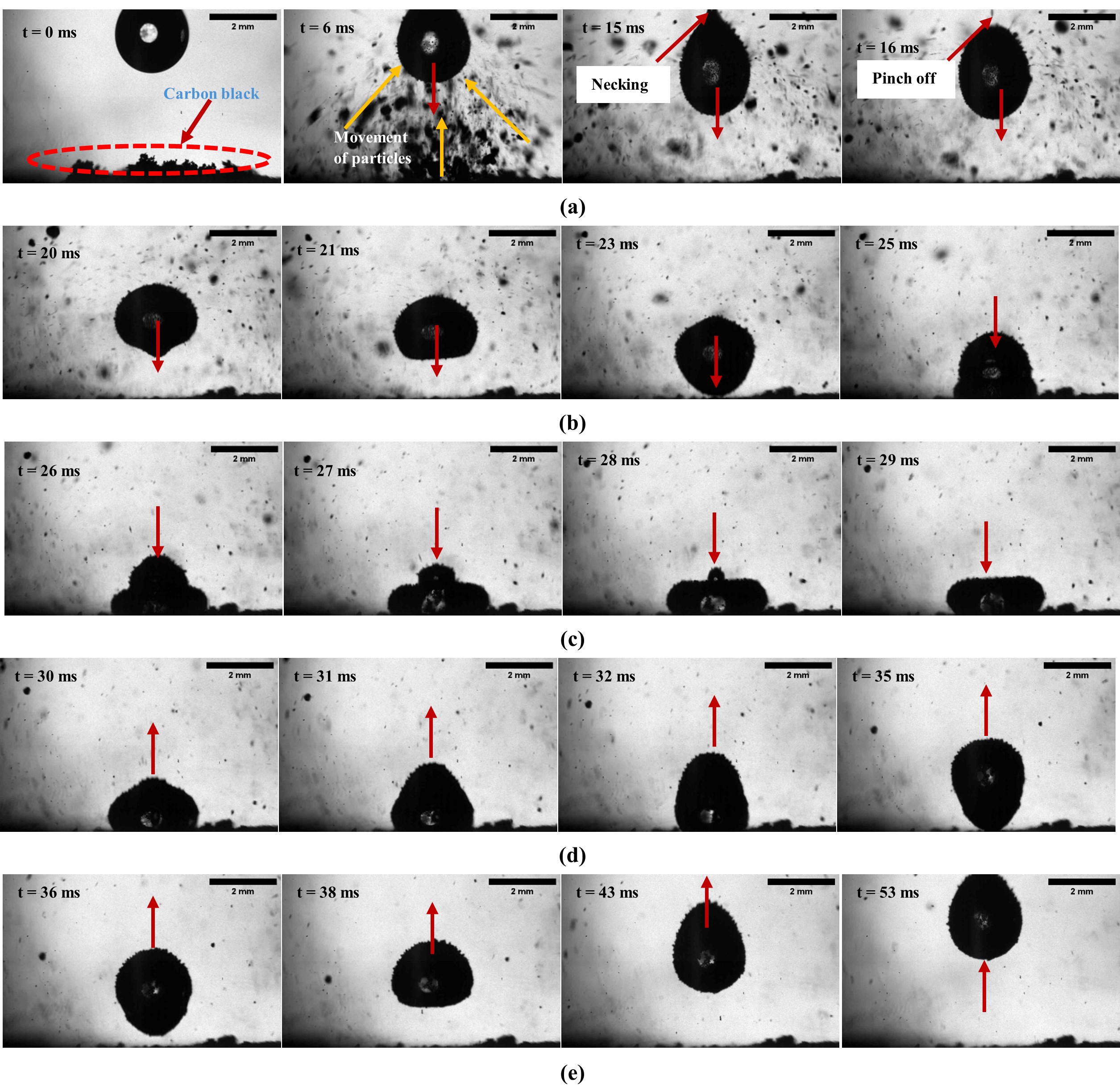}
     \caption{Time sequential images showing the dynamics of drop on the substrate covered with black carbon particles at 3.5 kV applied potential: (a): Drop elongation and pinch-off, (b): impact, (c): spreading, (d): recoiling, and (E): bouncing. The corresponding video is available as video 2 in SI {\color{blue}\href{https://drive.google.com/file/d/1AQxNhZay4o2W8yTPICjXL3pBLLIgePwX/view?usp=drive_link}{video2}}}
    \label{fig:3}
    \end{figure*}

In this experiment, a ridge of carbon particles is dispersed over the copper ground electrode, which is indicated in Figure \ref{fig:3} at t = 0 ms. The applied potential is varied from 2 kV to 4 kV at a step size of 0.5 kV. Figure  \ref{fig:3} shows the behavior of the droplet dynamics as well as particle migration under the application of the non-uniform electric field at 3.5 kV applied potential. It is observed that at a lower applied potential (2 kV), the droplet fails to detach from the capillary. However, at higher potentials (2.5 kV, 3 kV, and 3.5 kV), the droplet becomes unstable due to Rayleigh instability \cite{kull1991theory}. The charged droplet detaches from the capillary tip and descends onto the particle-covered ground electrode while capturing the particles. The expression of Rayleigh's charge is given in equation \ref{rayleigh_limit}.
\begin{equation}
    q_R = 8\pi\sqrt{\epsilon_0\gamma (D_{eq}/2)^3}
    \label{rayleigh_limit}
\end{equation}
Here $q_R$ is the Rayleigh charge on the droplet, $\epsilon_0$ is the permittivity of free space, $\gamma$, and $D_{eq}$ is the surface tension and the equivalent diameter of the droplet respectively.\\
Above 2.5 kV, the highly conductive carbon black particles and a droplet get charged.
In this study, the stable pendant droplet is generated before the application of the electric field (see Figure \ref{fig:3} at t = 0 ms). As soon as the electric field is applied, both droplet as well as particles get charges through Faradaic induction \cite{xu2018electron,biswal2023study}.
When the voltage is applied between the capillary and ground electrode covered with conducting particles, the electric field is established between the electrodes. This external electric field polarizes the particles and induces the dipole in the particles. This causes a charge separation between the particles. For example, if the direction of the electric field is toward the ground plate, negative charges will accumulate on the side of the particle facing the pin electrode. The positive charges will gather on the side facing the ground plate. Over a short period, determined by the particles' charge relaxation time, the positive charges on the particles will be neutralized by the negative charges on the ground plate induced by the same electric field. This results in a net negative charge on the particle, a phenomenon known as Faradaic induction \cite{biswal2023study}.\\
The carbon black particles, owing to their high conductivity, exhibit a minimal charging time off (charge relaxation time, $\tau = \epsilon/\sigma$ = 0.5 ps \cite{haus1989electromagnetic,biswal2023study}), leading them to migrate toward the charged droplet. It is observed that at an applied potential of 3.5 kV, the particles easily migrate and adhere to the droplet surface \cite{biswal2023study}, forming an encapsulated structure on the circumference of the charged droplet as shown in Figure \ref{fig:3} at t = 6 ms. During this period, the charged droplet continues to deform leading to the formation of a prolate shape (Figure \ref{fig:3} at t = 15 ms).\\
Subsequently, the droplet pinches off from the capillary at t = 16 ms and moves in a direction opposite to the particles, which is shown in Figure \ref{fig:3} at t = 16 ms). As the particles and the charged droplet interact with each other due to the electrostatic force of attraction, the particles adhere to the droplet's surface, rendering this partially hydrophobic. Once the droplet's surface becomes hydrophobic and impacts onto the metallic copper plate (Figure \ref{fig:3} at t = 23 ms). Then the charged droplet reverses its charge due to the method of charge induction. During the impact upon the copper plate, the charged droplet forms the shape of a pancake as shown in Figure \ref{fig:3} at t = 26, 27, and 28 ms. The maximum spreading of the droplet is observed at t = 29 ms as shown in Figure \ref{fig:3}. However, due to the high surface energy and the attractive force between the capillary and the copper plate, the charged droplet starts recoil at t = 30 ms and stands on the copper plate with minimum contact at t = 35 ms. After recoiling, the droplet is attracted toward the charged capillary (\cite{lee2015two,yao2018experimental}). The charged droplet again touches the tip of the capillary as shown in \ref{fig:3} at t = 53 ms. This back-and-forth motion of the droplet between the capillary and the ground electrode continues. It is observed that the charged droplet after capturing the carbon black particle makes five times back-and-forth motion between the capillary and copper plate.  The droplet moves away from the target surface and goes out of the frame after t = 260 ms. For more clarity, the reader is advised to see video 2 in SI.

\subsubsection{\label{sec:3.3}Droplet dynamics impinging on low conducting and hydrophobic particles (silane-coated glass beads)}
\begin{figure*}
    \centering
    \includegraphics[width=1\linewidth]{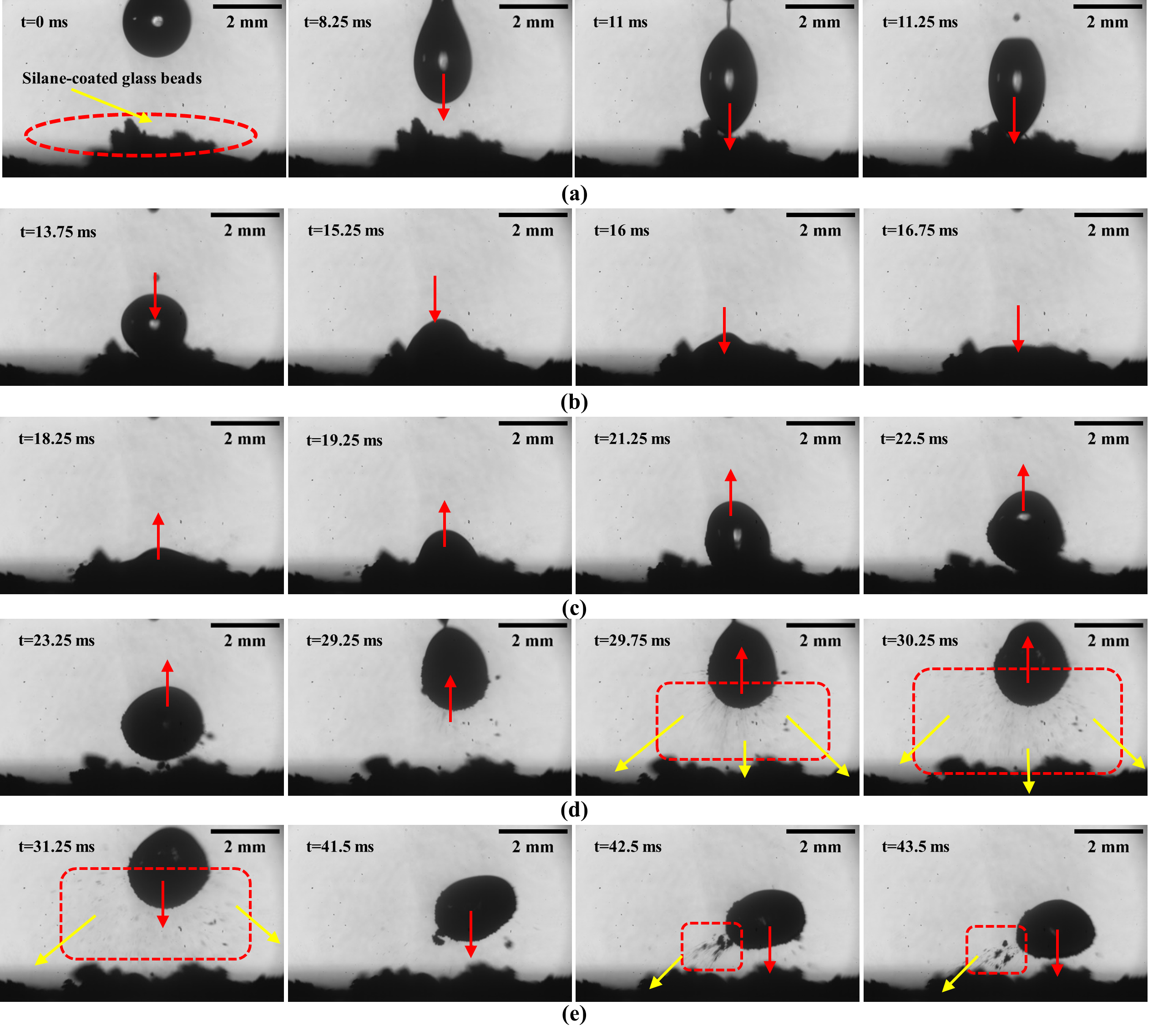}
    \caption{Time sequential images showing the dynamics of drop on the substrate covered with silane coated particle at 3.5 kV applied potential: (A) pinch-off and impact, (B) spreading, (C) recoiling, (D) bouncing towards the capillary, and (E) again falling on the substrate. The corresponding video is available as video 3 in SI {\color{blue}\href{https://drive.google.com/file/d/1aBa5NABabrHsKUhsL09KWZ4MWwSWB297/view?usp=drive_link}video3}}
    \label{fig:9}
\end{figure*}

In this case, the droplet impacts the silane-coated glass beads dispersed on the ground electrode. The dynamics of the droplet at an applied potential of 3.5 kV at different time instants is shown in Figure \ref{fig:9}.  
The droplet remains attached to the capillary at t = 0 ms as indicated in Figure \ref{fig:9} without any application of electric field. Once the electric field is applied, the droplet gets charged due to the induction charging mechanism. The charged droplet undergoes capillary stretching and forms along necking with the capillary tip at a t = 8.25 ms and t = 11 ms respectively as depicted in Figure \ref{fig:9}. It is important to note here that the particles do not move toward the charged droplet due to its poor electrical conductivity. So, no migration of the particles means that they do not acquire any charge at 3.5 kV compared to carbon black particles in section \ref{sec:3.2}.\\

Then the droplet pinches off from the capillary tip and touches the particle surface as shown in Figure \ref{fig:9} at t = 11.25 ms. When the charged droplet falls on the ridge of the particles, it contacts the copper plate by breaking the particle ridge as shown in Figure \ref{fig:9} at t = 13.75 ms, 15.25 ms, and 16 ms. During this stage, the droplet deforms in the radial direction ($D_h$) while minimizing along the vertical direction. ($D_v$) as shown in Figure \ref{fig:9} at t = 16.75 ms. The phenomenon of drop spreading on the particle surface results in the conversion of kinetic energy to the surface energy of the droplet \cite{tian2022does}.

 When hydrophobic silane-coated particles adhere to the droplet surface, the droplet surface partially turns hydrophobic, similar to carbon black particles. Due to the hydrophobic nature of the droplet, the droplet recoils at t = 18.25 ms and makes minimum contact area with the particle surface at t = 22.5 ms.  During this recoil phenomenon of the droplet, high surface energy is converted into kinetic energy \cite{deng2023behaviors}. Whereas, the droplet reverses its polarity upon impact on the ground electrode. Then the droplet moves toward the highly charged capillary due to electrostatic force of attraction between the capillary and the droplet.

The droplet changes shape from prolate to oblate and vice versa as it moves towards the capillary. This shape change is due to competition between non-uniform electric stress, surface tension, and gravity force. The droplet polarity reverses as it contacts the capillary. At this time, the attached particles on the droplet circumference are repelled and thrown away along the radial direction (indicated with the yellow colored arrow) due to droplet deformation and repulsive electrostatic force as shown in Figure \ref{fig:9} at t = 29.75 ms and 30.5 ms.\\
The charged droplet is drawn back towards the ground electrode by gravity and electrostatic attraction. During the downward motion of the charged droplet, some of the remaining particles are again radially thrown away (indicated with the yellow-colored arrow) by the droplet as shown in Figure \ref{fig:9} at t = 31.25 ms. It is also observed that when the droplet is about to make contact with the particle surface, some of the particle clusters are repelled away from the target surface as shown in Figure \ref{fig:9} at t = 42.5 ms and 43.5 ms.\\
The charged droplet repeats this process two times back-and-forth motion between the capillary and copper plate while removing the particles. Finally, at t = 167.5 ms, the droplet leaves the frame (For more clarity, the reader is advised to see video 3 in SI).

In this case, the particles are captured by the adhesive forces between the droplet and the particles. The particles are removed from the target substrate due to the droplet's continuous shape changes and repulsive electric force.\\\\
\begin{figure*}
    \centering
    \includegraphics[width=1\linewidth]{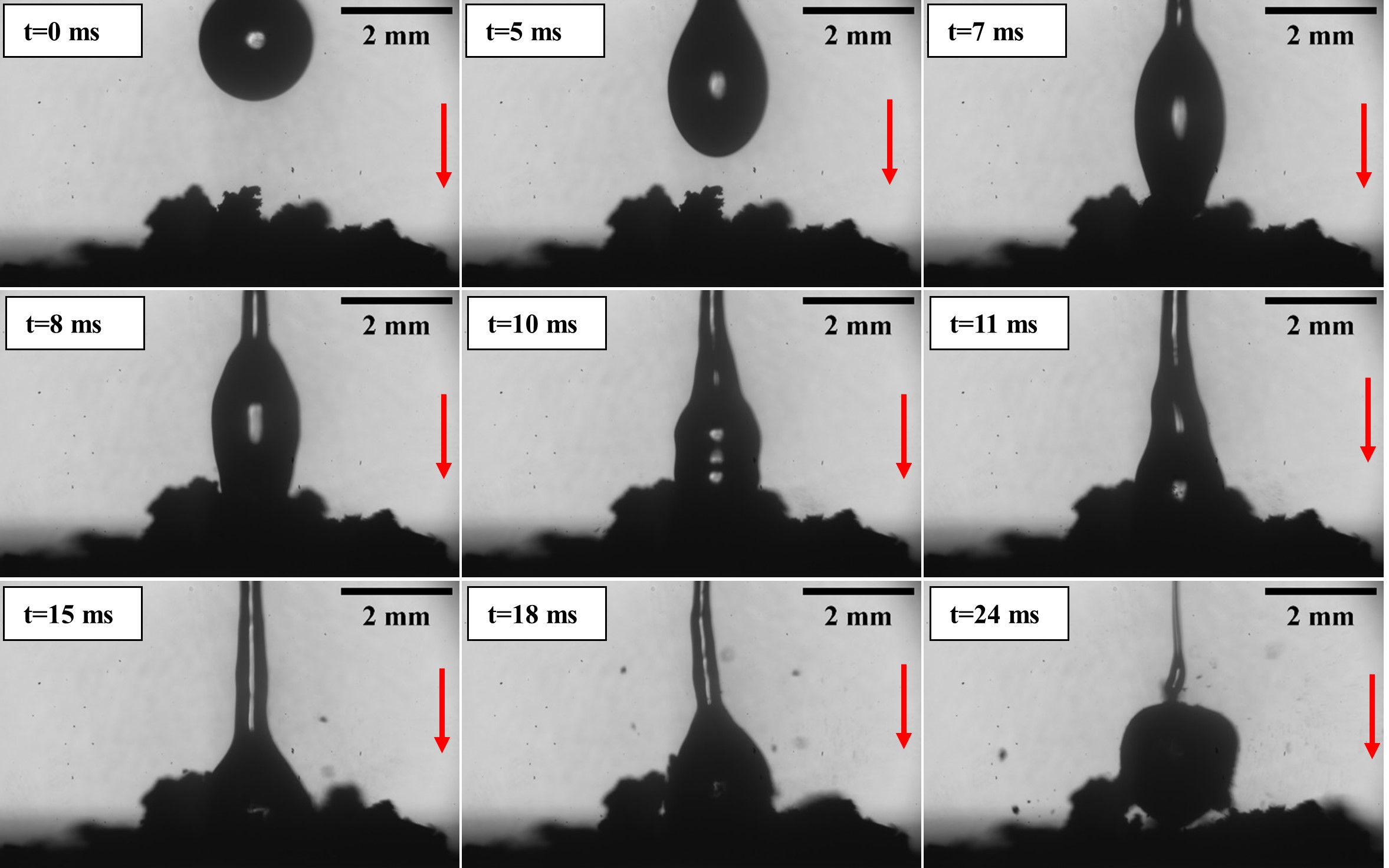}
    \caption{Time sequential images showing dynamics of drop impinging on a substrate covered with silane coated particle moving at 4 kV applied potential. The corresponding video is available as video 4 in SI{\color{blue}\href{https://drive.google.com/file/d/1ZVy7imMO08_-W4xvcZnoZ7LEkI_vshwX/view?usp=drive_link}{video4}}}
    \label{fig:10}
\end{figure*}
From the above observations, the silane-coated glass beads fail to migrate toward the charged droplet at 3 kV. To ensure the possibility of charging silane-coated glass beads at higher potential through charge induction. The same experiment is conducted by varying the applied potential from 3.5 kV to 4 kV. It is observed that at 4 kV, an electric breakdown occurs between the electrodes. This electric breakdown results in making a bridge between the droplet and particle surface as shown in Figure \ref{fig:10} at a time t = 10 ms without undergoing breaking, and the necking formation. With this high applied potential, the silane-coated particles do not show upward movement. This stagnation behavior of the silane-coated particles ensures either a lack of charge transformation to the particles or a requirement of high charging time.
\subsubsection{\label{sec:3.4}Droplet dynamics impinging on hollow glass beads}
Another set of experiments is conducted to explore the dynamic implications of charged droplets falling on hydrophilic particles, specifically hollow glass beads. These glass beads exhibit favorable wetting properties with a contact angle of 23$^\circ$ and possess moderate electrical conductivity.\\ 

\begin{figure*}
    \centering
    \includegraphics[width=1\linewidth]{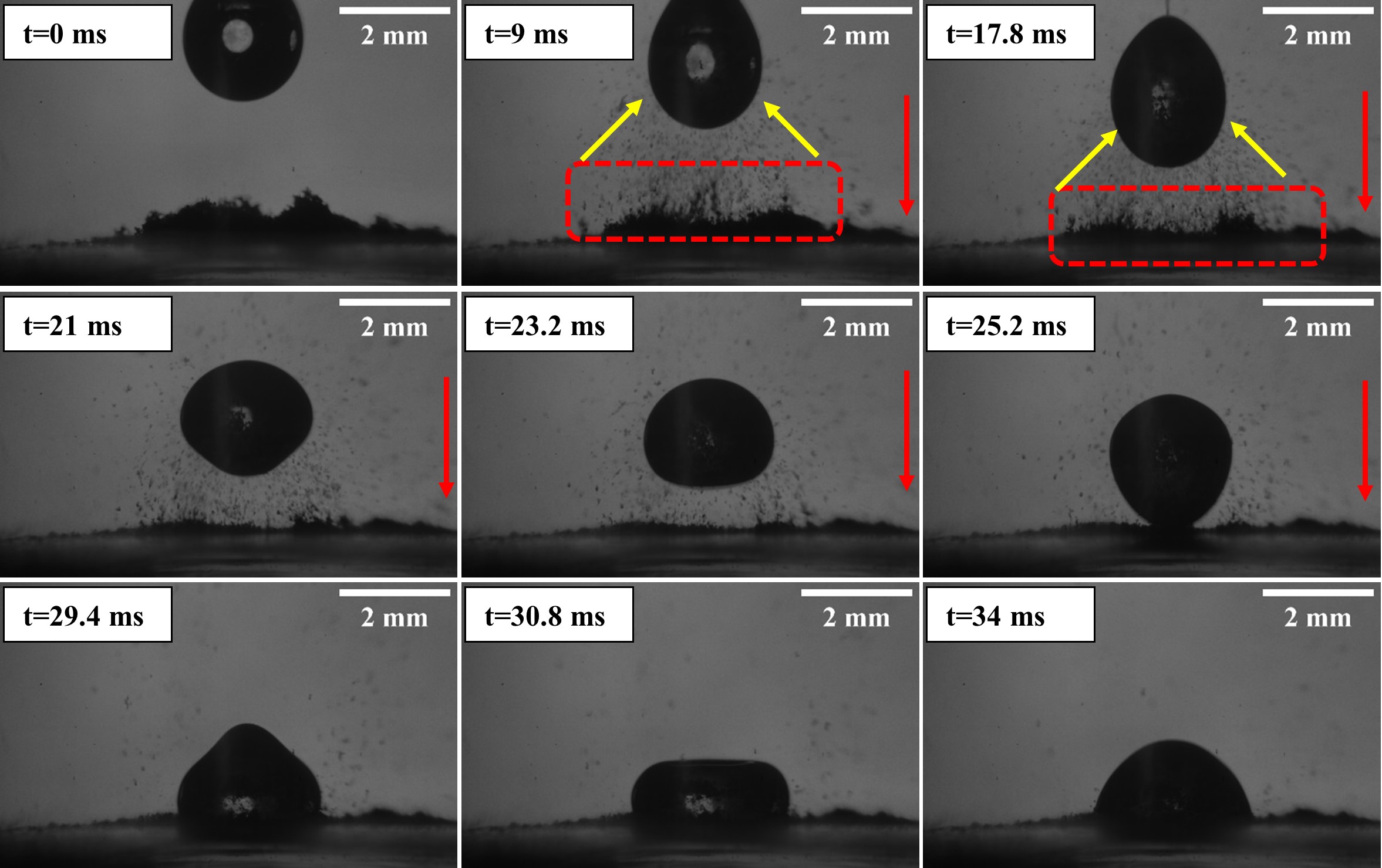}
    \caption{Time sequential images showing the dynamics of a drop impinging on the substrate covered with hollow glass beads moving at 3.5 kV applied potential. {\color{blue}\href{https://drive.google.com/file/d/1b8mWVcNCqWDKbQbYRFalqXSukuOMeCj5/view?usp=drive_link}{video5}}}
    \label{fig:11p}
\end{figure*}

When the electric field is applied, the particles get charged and start to migrate towards the charged droplet similar to section \ref{sec:3.2}, as shown in Figure \ref{fig:11p} at t = 9 ms. During this period, the electric force dominates over the capillary force resulting in the droplet pinch-off as shown in Figure \ref{fig:11p} at t = 17.8 ms. During this droplet elongation and pinch-off, the particles are simultaneously captured by the charged droplet due to the electrostatic force of attraction. Then droplet starts moving toward the copper plate and touches the surface at t = 25.2 ms. It is observed that due to the hydrophilic nature of the particles, the particles adhere to the droplet and incorporate inside the droplet without attaching to its surface \cite{biswal2023study}. This behavior is different, with no particles adhering to the droplet's surface. This ensures that the droplet surface remains hydrophilic. Due to low surface energy, the droplet fails to recoil on the particle surface. However, this ensures that only electric force is enough to pull up the droplet toward the charged capillary. Therefore, the droplet remains intact with the surface upon impact after capturing hydrophilic particles. This observation indicates that although the charged droplet attracts particles through electrostatic forces, it does not exhibit the typical rebounding behavior.

\subsection{\label{sec:3.5}Discussions}
\begin{figure*}
    \centering
    \includegraphics[width=0.8\linewidth]{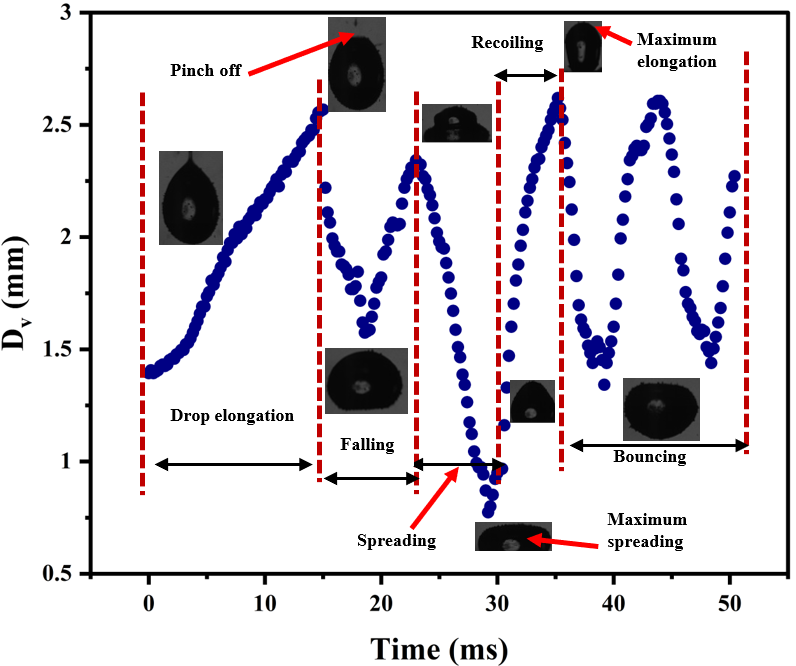}
    \caption{Different regimes of drop impact on the surface covered with the carbon particles at 3.5 kV applied potential.}
    \label{fig:11}
\end{figure*}

The previous section \ref{sec:3:Results and discussions} successfully provided the experimental observations of charged droplet dynamics and particle elimination methods. The observation shows the droplet becomes unstable and detaches from the capillary tip when the electric force surpasses the capillary force. This process mostly involves drop deformation, pinch-off, impact, particle collection, recoiling, bouncing back, and so on.  Figure \ref{fig:11} shows the change in the aspect ratio of the droplet during this process, where, $D_h$ and $D_V$ are the major axis diameter and minor axis diameter, respectively. $D_V$ gradually increases with the application of the electric field and the pinch-off occurs possibly due to the Rayleigh instability (see Figure \ref{fig:11}). Thereafter, the droplet travels toward the capillary with changing the droplet shape from oblate to prolate and vice versa. $D_V$ becomes minimum upon impact and leads to the drop into a flattening position. During the recoiling phenomenon, the $D_V$ gradually increases and attains the maximum value as shown in Figure\ref{fig:11}. This process repeats several times until the droplet leaves the frame.\\ 
The equivalent diameter of the droplet is obtained by assuming a spheroid using the empirical relation \ref{eq:6} \cite{deng2023behaviors}.
\begin{equation}
    D_{eq} = (D_vD_h^2)^{1/3}
    \label{eq:6}
\end{equation}
Where $D_{eq}$ is the equivalent diameter. The maximum equivalent diameter of the falling droplet at 3.5 kV is obtained equal to 2.3 mm. \\\\
In classical drop dynamics, the variation of drop shape is measured through non-dimensional numbers: Reynolds number (Re), Weber number (We), and Ohnesorge number (Oh). The Reynolds number measures the prevalence of inertia force in comparison to viscous force, while the Weber number assesses the significance of inertial force relative to surface tension \cite{yarin2006drop}. Ohnesorge number measures the relative importance of viscous force, surface tension force and inertial force \cite{fakhari2009simulation}. The mathematical expressions for these non-dimensional parameters are provided in Eqs. \ref{eq:7}, \ref{eq:8}, and \ref{eq:9}. 

\begin{equation}
    Re = \frac{\rho vD_{eq}}{\mu_a}
    \label{eq:7}
\end{equation}
\begin{equation}
    We = \frac{\rho D_{eq} v^2}{2 \gamma}
    \label{eq:8}
\end{equation}
\begin{equation}
    Oh = \frac{\mu_a}{\sqrt{\frac{\rho \gamma D_{eq}}{2} }}
    \label{eq:9}
\end{equation}
Here, $\rho, \gamma, v, D_{eq}$, are the density, interfacial tension, impact velocity, and equivalent diameter of the droplet respectively. $\mu_a$ is the dynamic viscosity of the suspending medium (air). \\
The maximum Reynolds number, Weber number, Ohnesorge number, and impact velocity are 1190, 7, 0.0032, and 0.49 m/s respectively. As a result of the combination of a low Weber number and impact speed, the droplet remained intact without breaking up into the daughter droplets during the impact.\\
In the case of the charged droplet, the shape deformation of the droplet is explained using the non-dimensional electric capillary number provided in equations \ref{eq:10} \cite{kim2007deformation} .\\
\begin{equation}
    Ca_E = \frac{\epsilon_0E^2 D_{eq}}{2 \gamma}
    \label{eq:10}
\end{equation}
Where $\epsilon_0$ and $E$ are permittivities of free space, and local electric field strength, respectively.
\subsubsection{\label{sec:3.1}Calculation of charge on the droplet}
\begin{figure}
    \centering
    \includegraphics[width=1\linewidth]{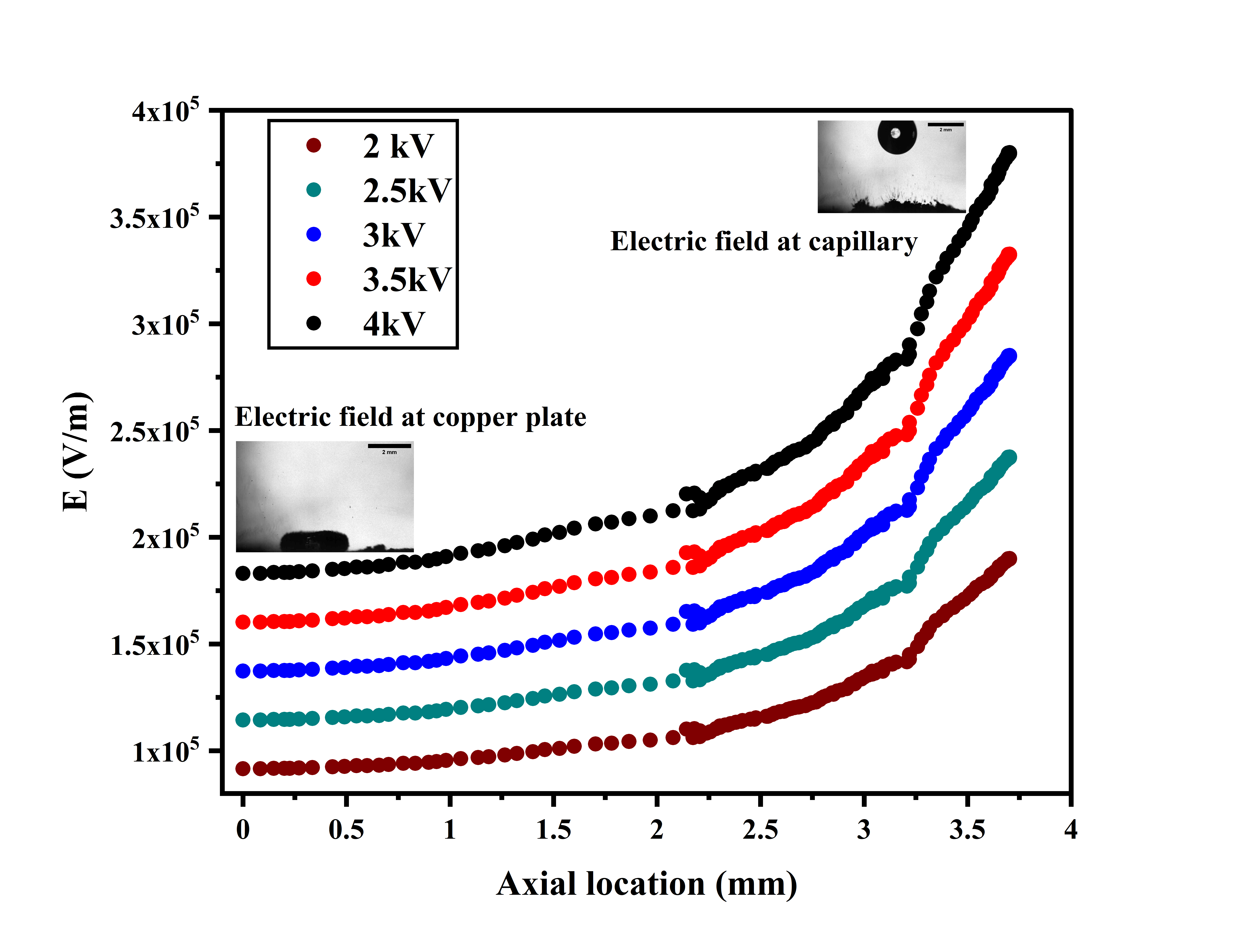}
    \caption{Electric field distribution of pin-plate electrode configuration along the axial direction with fixed electrode separation distance, 6 mm. }
    \label{fig:electric_field}
\end{figure}

In this study, the droplet and the particles are exposed to the non-uniform DC electric field, in which the strength of the electric field is the function of the axial and radial distance between the pin-plate electrode. The variation of the electric field distribution and charge on droplets are obtained by solving the sphere–plate expression proposed by Fernando. For sphere–plane electrode configuration, the expression for the electric field distribution along axial and radial direction \cite{dall2009solution, biswal2023study} are expressed in Equations \ref{equation:1} and \ref{equation:2}. 
\begin{equation}
E_y(r,y) = (D_{eq}/2) V \sum_{i=0}^{\infty} \left[\frac{\xi_i (y-z_i)}{[(y-z_i)^2+r^2]^{2/3}} - \frac{\xi_i (y+z_i)}{[(y+z_i)^2+r^2]^{2/3}}\right]
\label{equation:1}
\end{equation}
\begin{equation}
E_r(r,y) = (D_{eq}/2) V \sum_{i=0}^{\infty} \left[\frac{\xi_i}{[(y-z_i)^2+r^2]^{2/3}} - \frac{\xi_i}{[(y+z_i)^2+r^2]^{2/3}}\right]
\label{equation:2}
\end{equation}
Here $D_{eq}$ is the equivalent diameter of the droplet, $V$ is the applied potential, $z_i$, and $\xi_i$ are the axial position and normalized charge of the $i^th$  image charge given by recurrent relations.
\begin{equation}
z_i = \frac{(D_{eq}/2)^2}{z_0+z_{i-1}}
\label{equation:3} 
\end{equation}
\begin{equation}
\xi_i = \frac{q_i}{q_0}
\label{equation:4} 
\end{equation}
\begin{equation}
q_i = \frac{(D_{eq}/2) }{z_0+z_{i-1}}  q_{i-1}
\label{equation:5} 
\end{equation}
A MATLAB program was developed using the above equations to compute the electric field for a droplet-ground (sphere-plane) electrode configuration. The electric field distribution for sphere-plane electrode configuration at 2, 2.5, 3, 3.5, and 4 kV is determined and shown in Figure \ref{fig:electric_field}. The corresponding maximum charge of the droplet at the capillary and copper plate with 6 mm separation is evaluated as given in Table \ref{tab:charge}.
\begin{table*}
\caption{\label{tab:charge} Charge of the droplet in non-uniform electric field distribution: q1: charge of the droplet at capillary and q2: charge of the droplet at copper plate}
\begin{ruledtabular}
\begin{tabular}{ccdd}
Voltage (kV) & q1 (nC) & q2 (nC)\\
\hline
 2 & 2.52 & -0.96 \\
2.5 & 3.15 & -1.2 \\
 3 & 3.78 & -1.44\\
3.5 & 4.41 & -1.68\\
 4 & 5.04 & -1.92\\
\end{tabular}
\end{ruledtabular}
\end{table*}

\subsubsection{Force analysis on charged droplet}
\begin{figure}
    \centering
    \includegraphics[width=1\linewidth]{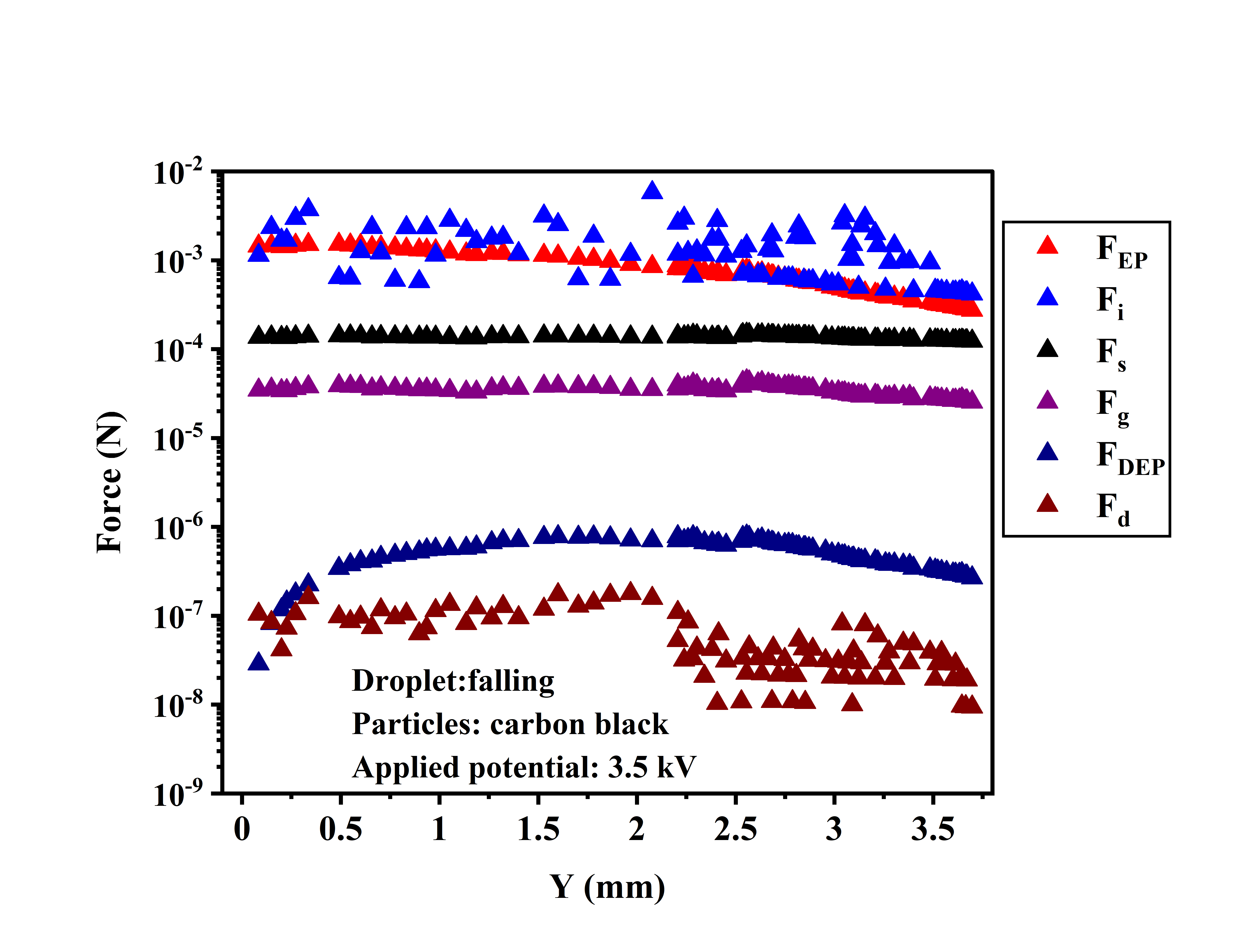}
    \caption{The variation of the important forces as a function of distance Y acting on a charged droplet at 3.5 kV applied potential. Where Y is the distance from the particle surface to the tip of the capillary.}
    \label{fig:12}
\end{figure}

Table \ref{tab:table2} shows the expression of primary forces affecting droplet dynamics in a non-uniform electric field \cite{roux2008forces}. The magnitudes of these forces are presented as a function of the distance between the tip of the capillary to the surface of the ground electrode (Y (mm)) in Figure \ref{fig:12}. The analysis reveals that the electrophoretic (EP) force exerts the greatest influence on the droplet, with a maximum magnitude of approximately $10^{-3}$ N. In contrast, the dielectrophoretic (DEP) force has minimum influence on the droplet on the order of $10^{-7}$ N, as shown in Figure \ref{fig:12}. The DEP has no significant effect in the current study.\\
The Bond number ($Bo = \rho ga^2/\gamma$), which is the ratio of gravitational force to surface tension \cite{liu2014numerical, birbarah2015comprehensive} of the uncharged droplet is in the range of  0.09 to 0.13. This shows that the droplet shape of an uncharged droplet is dominated by the surface tension

When an electric field is imposed, the droplet acquires a charge, and an electric force is exerted on it. The electric force is on the order of 10 times greater than the surface tension force as shown in Figure  \ref{fig:12}. Whereas, the electric capillary number is determined in the range from 0.08 to 0.13 respectively. The larger value of the electric capillary number shows a greater contribution to the drop deformation in the presence of the electric field. In conclusion, the electric force shows dominant behavior over the surface tension force in the presence of the electric field.
\begin{table}
\caption{\label{tab:table2} key forces acting on the droplet }
\begin{ruledtabular}
\begin{tabular}{ccddd}
Type of Forces & Expressions \\
\hline
Coulombic/electric force  & $F_{EP}$ = qE  \\
DEP & $F_{DEP} = \frac{\epsilon_d - \epsilon_m}{\epsilon_d + 2\epsilon_m}2\epsilon_m\pi (D_{eq})^3 \nabla E^2$ \\
Gravity force  & $F_g=  \frac{4}{3}\pi (D_{eq})^3\rho g$\\
Stokes drag force & $F_d=  3\pi \eta_a D_{eq}(v-u_a)$   \\
Inertia force & $F_i = ma$ \\
Surface tension force & $F_s = \gamma D_{eq} $ \\
\end{tabular}
\end{ruledtabular}
\end{table}
\subsection{Drop deformation}
\begin{figure}
    \centering
    \includegraphics[width=1\linewidth]{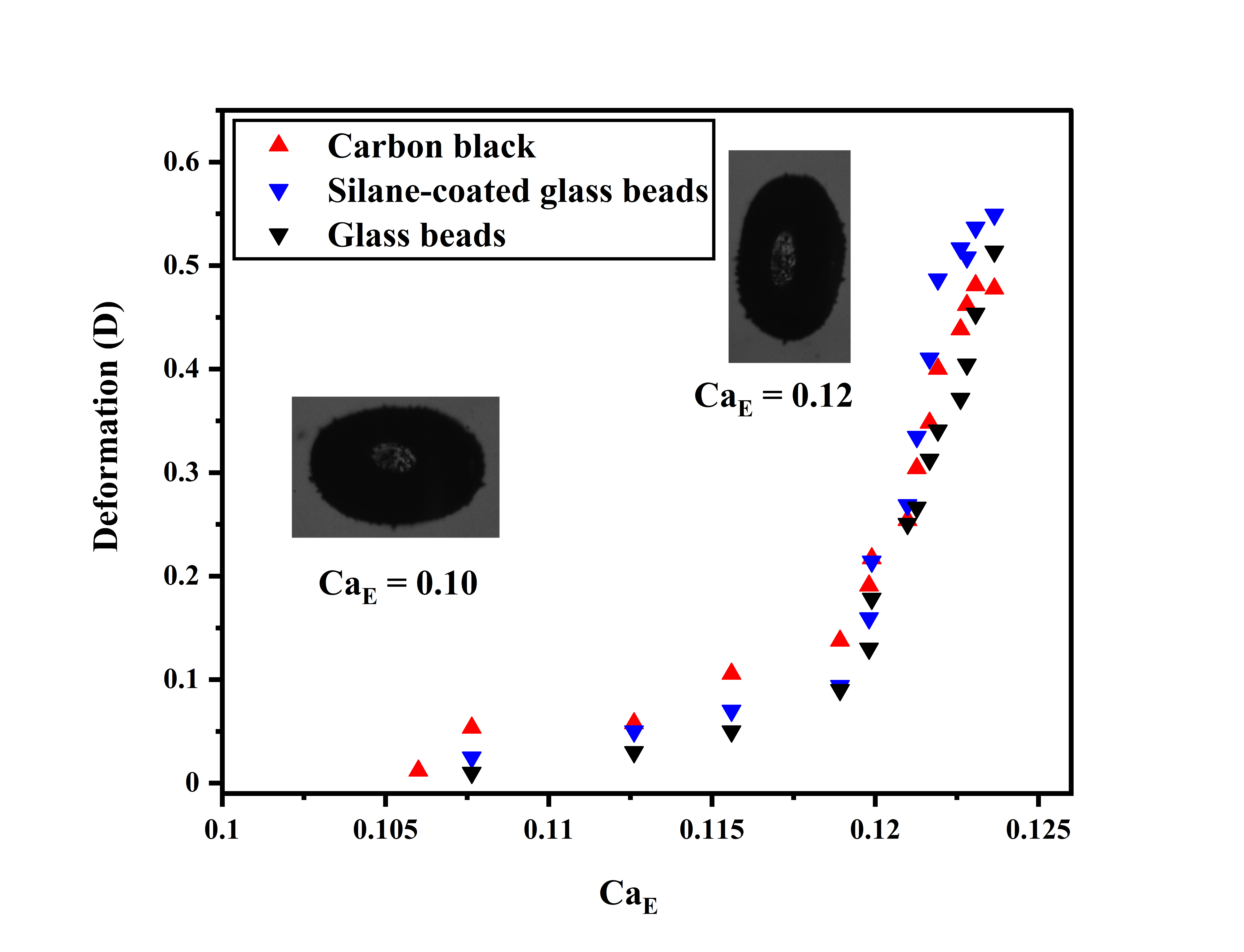}
     \caption{Drop deformation as a function of capillary number at applied potential 3.5 kV.}
    \label{fig:drop_deformation}
\end{figure}

The drop deformation strongly depends on the electric capillary number \cite{naz2023three}. The degree of deformation (D) is defined as \cite{naz2023three, nudurupati2008concentrating}
\begin{equation}
    D = \frac{D_h - D_v}{D_h + D_v}
\end{equation}
When the particles get incorporated on the droplet surface, the surface phenomenon of the droplet changes. Due to the particle encapsulation and the non-uniform electric field, the drop deforms continuously. The effect of the properties of three different particles on drop deformation is elucidated in Figure \ref{fig:drop_deformation} at 3 kV. This figure shows that the droplet deforms from oblate to prolate under a non-uniform electric field with particle encapsulation. The magnitude of the drop deformation increases with increasing the capillary number with no significant change in the deformation factor up to the capillary number up to 0.115. It is interesting to see that the data collapses very well indicating that the transformation from the oblate to prolate occurs above the capillary number 0.115 (shown in Figure \ref{fig:drop_deformation}). Therefore, it confirms that the particle properties (wettability) and its encapsulation on the drop surface have no significant role in the drop deformation.
\subsubsection{\label{sec:3.6}Different regimes of the droplet during impact under the influence of electric field}
\begin{figure*}
    \centering
    \includegraphics[width=1\linewidth]{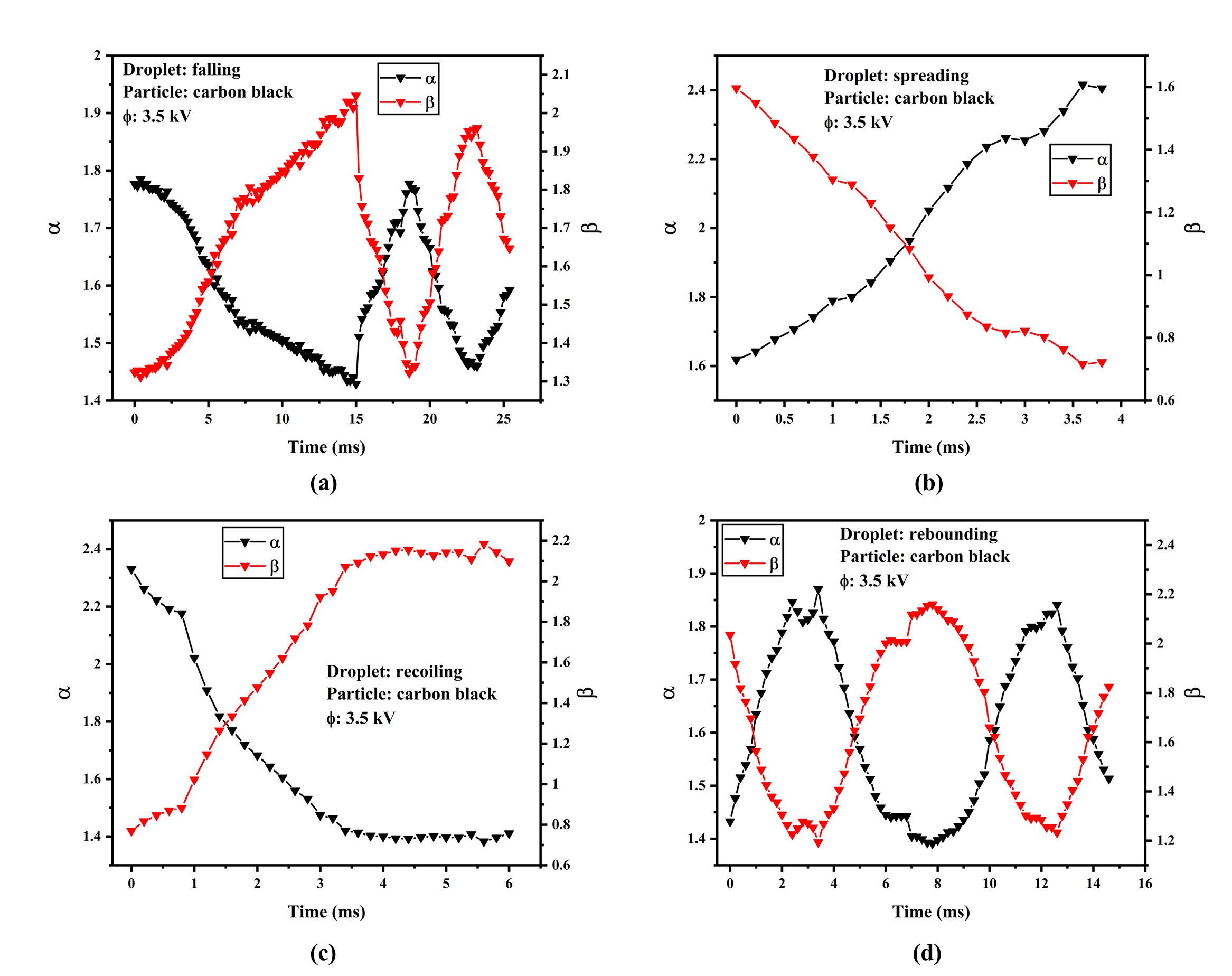}
    \caption{Effect of spreading factors ($\alpha$ and $\beta$) under non-uniform electric field at 3.5 kV: (a) falling, (b) spreading, (c) recoiling, and (d) bouncing}
    \label{fig:drop_dynamics}
\end{figure*}

The droplet dynamics under the application of an electric field exhibit four different regimes and illustrated in Figure \ref{fig:drop_dynamics}: (a) shape deformation during stretching and traveling towards the substrate, (b) spreading in the radial direction, (c) The recoiling of the droplet during impact on the surface of particles and (d) bouncing. The shape deformation of the droplet can be described using non-dimensional parameters (shape deformation factors $\alpha$ and $\beta$), which are defined as $\alpha = D_h/D_{eq}$ and $\beta = D_v/D_{eq}$ \cite{crooks2001role, pasandideh1996capillary, zhang1997dynamic}.
Figure \ref{fig:drop_dynamics} (a) shows the change in the shape of the droplet during falling (after detaching from the capillary tip) to before making contact with the surface of particles. The shape deformation factors $\alpha$ and $\beta$ change with time, which indicates that the continuous changes in shape during falling. The maximum values of $\alpha$ and $\beta$ are 1.75 and 2, respectively. The droplet starts spreading in the radial direction with increasing of $\alpha$ as soon as it contacts to the particles dispersed on the surface (see Figure \ref{fig:drop_dynamics} (b)).  
The corresponding maximum and minimum deformation factors ($\alpha$ and $\beta$) are determined to be 1.6 and 0.8 respectively.\\\\

Thereafter, the $\alpha$ starts to decrease with the corresponding increase of $\beta$, which indicates the reduction of the major axis and increase of the minor axis of the droplet which means the droplet changes its shape from flattened to prolate shape. This phenomena is termed as recoiling and presented in Figure \ref{fig:drop_dynamics} (c).


This retraction of the droplet takes place because surface energy gradually converts into kinetic and eventually into potential energy. The corresponding minimum and maximum values of $\alpha$ and $\beta$ are 1.4 and 2.4 respectively as shown in Figure \ref{fig:drop_dynamics} (c).
\\


Figure \ref{fig:drop_dynamics} (d) shows the variation of $\alpha$ and $\beta$ as a function of time during the rebounding of the droplet. The rebounding phenomenon takes place when the droplet travels from the particle surface to the capillary tip.
The rebounding of the droplet occurs because of the relative change of electric stress ($P_E= \frac{1}{2} \epsilon_0 \epsilon_a E^2$), capillary stress ($P_\gamma= 2\gamma⁄a$), gravity effect ($P_g= \rho gh$), and pressure dynamics ($P_d= \frac{1}{2} \rho v^2$) \cite{deng2023behaviors}.  Where $\epsilon_0$ and $\epsilon_a$ are the permittivities of free space and relative permittivity of air respectively. $E$, $\gamma$, $\rho$, and $v$ are the local electric field strength, surface tension, density, and velocity of the droplet respectively. $g$ and $h$ are acceleration due to gravity and droplet height from the ground electrode, respectively.  Gravity effect and capillary stress attempt to impede droplet recoiling and rebounding, but when electric stress surpasses them, the droplet initiates rebounding as shown in Figure \ref{fig:drop_deformation}. The same physics was also discussed by \citeauthor{deng2023behaviors}.\\
During the rebounding process, the gradual variation of the electric field gradient causes the modification of electric stress in the vertical direction (from the surface of the particle to the capillary tip).The variation of electric stress causes the droplet to alter its shape from prolate to oblate and vice versa. 
\\
\subsubsection{Effect of electric field on droplet dynamics}
\begin{figure*}
    \centering
    \includegraphics[width=1\linewidth]{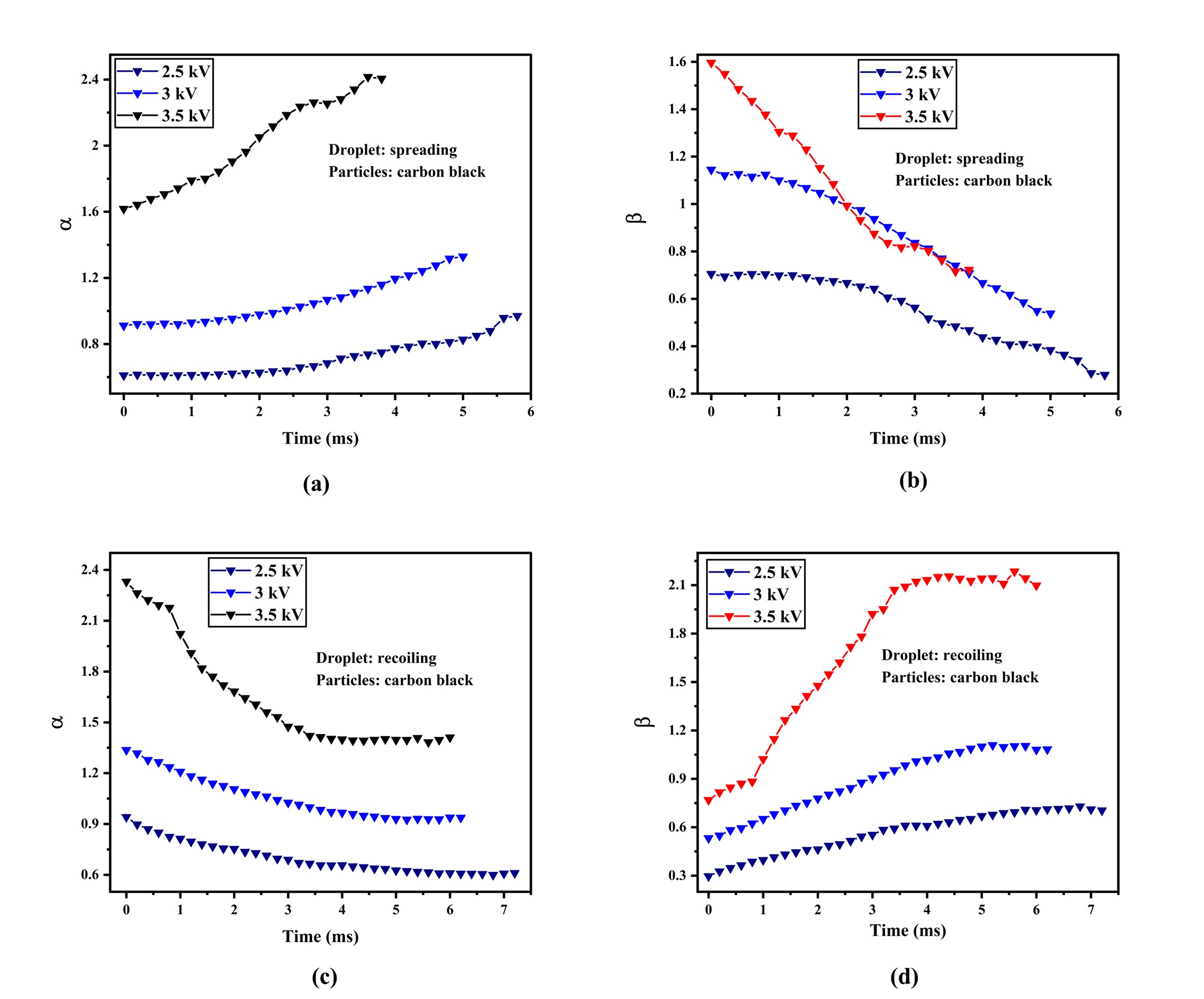}
     \caption{Effect of electric field on $\alpha$ and $\beta$ during spreading and recoiling of the droplet.}
    \label{fig:effect_of_electric_field}
\end{figure*}

 The experiments have been also done for different applied potentials 2, 2.5, 3, 3.5, and 4 kV. High-speed videography reveals distinct outcomes at different potentials. At 2 kV, there is no droplet breakage or observable dynamics. At 2.5, 3, and 3.5 kV, with a droplet diameter of approximately 2 mm, four regimes: impact, spreading, recoiling, and bouncing, are observed. Beyond 4 kV, Rayleigh instability induces electric breakdown, leading to droplet bridging between electrodes. The electric field strength as well as charge on the droplet near the capillary tip, obtained through MATLAB code, are in the range from 1.8$\times$10$^5$ to 3.7$\times$10$^5$ V/m and 3.8$\times$10$^{-9}$ to 7.8$\times$10$^{-9}$ C, respectively. Simultaneously, the electric capillary number falls within the ranges of 0.08 to 0.13. Notably, at higher electric potentials, both spreading factors $\alpha$ and $\beta$ in the spreading and recoiling regimes exhibit higher values compared to lower potentials as depicted in Figure \ref{fig:effect_of_electric_field}. This signifies that the contact area between non-conducting particles (silane-coated glass beads) and the droplet is more at higher applied potentials, so more particles will be attached to the droplet surface. This shows that the particle removal capacity increases with an increase in applied potential.
\subsubsection{Effect of particle wettability on drop recoiling}
\begin{figure}
    \centering
    \includegraphics[width=1\linewidth]{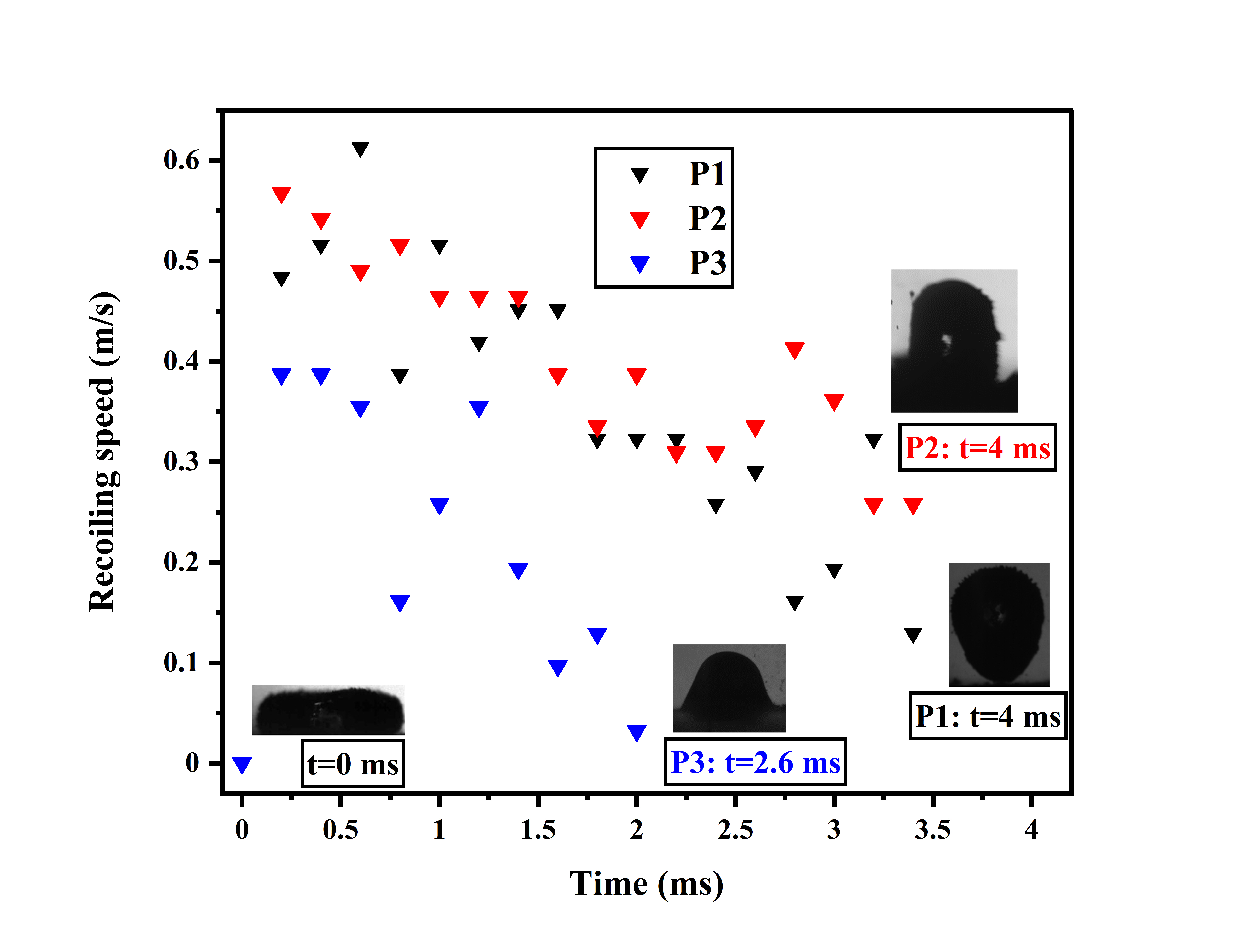}
    \caption{The velocity of the droplet at apex region (liquid-air interface) during recoiling phase at 3.5 kV. The velocity is measured by extracting the images frame by frame. Where P1, P2, and P3 denote hollow carbon black particles, silane-coated glass beads, and glass beads,  respectively.}
    \label{fig:recoiling_speed}
\end{figure}

The movement of the apex of the drop during the recoiling phase is quantified upon impact on three different particles. In the case of the drop impact on glass beads, the drop recoils with its apex maximum speed of 0.38 m/s. After a certain time, the apex speed of the drop gradually reduces and settles at a speed of 0.03 m/s. Hence the drop impact on the glass beads fails to recoil and rebound. In the case of drop impact on silane-coated glass beads, the drop starts recoiling with a maximum speed of 0.56 m/s and attains a minimum speed of 0.25 m/s. Whereas, the drop impact on carbon black recoils with a maximum speed of 0.48 m/s. After a certain time drop, it attains a minimum speed of 0.13 m/s before rebounding. For both carbon black and silane-coated particles, the drop maintains a certain speed and gets bounced back due to the electrostatic force of attraction. It is observed that the apex speed of the drop changes non-uniformly as shown in Figure \ref{fig:recoiling_speed}. It is also observed that the apex of the drop upon impact on silane-coated glass beads moves slightly higher compared to carbon black as shown in Figure \ref{fig:recoiling_speed}.
\section{Summary and Conclusions}
In this study, we experimentally examine the dynamic behavior of a millimeter-sized charged droplet upon impact on the particles dispersed on the ground electrode. The investigation utilizes a pin-plate electrode configuration and high-speed imaging technique to analyze the dynamic characteristics of the droplet with particle removal. Mathematical modeling is employed to determine the charge on the droplet and the non-uniform electric field distribution at different applied potentials (2, 2.5, 3, 3.5, and 4 kV). The vital findings of this study are delineated below.
\begin{itemize}
    \item The optimum applied potential that causes the drop pinch-off from the tip of the capillary is observed at an applied potential above 2 kV.
    \item When the electric field is applied between the electrodes. The particles acquire charge through induction due to their high electric conductivity. The particles migrate toward the charged droplet due to the electrostatic force of attraction. The particles get encapsulated on the droplet surface leading to the formation of the hydrophobic droplet surface.  The droplet with hydrophobic circumference makes continuous back-and-forth motion between capillary and ground electrodes during impact. The five-times back-and-forth motion has been observed. Thereafter, the droplet left the frame.
     \item The silane-coated glass beads do not migrate toward the charged droplet due to their bad conducting in nature even at a higher electric field (3.5 kV). Therefore, the particles get captured on the droplet surface during impact. This is due to adhesiveness between the droplet and the particles. In this case, the droplet surface partially achieves the hydrophobic nature. The droplet encapsulated with particles exhibits the rebounding phenomena during impacting on the surface. During the rebounding, the particles on the droplet surface are repelled and thrown away radially due to both continuous drop deformation and repulsive electrostatic force.
     \item In the case of glass beads (hydrophilic), the particles migrate and get captured inside the droplet due to the electrostatic force of attraction. Therefore, the droplet surface remains hydrophilic, hindering the rebounding of the drop upon impact.
     \item The charged droplet attains the maximum spreading at a higher applied potential. This signifies that more particles may be attached to the droplet surface upon impact and leads to cleaning more particles on the surface.
     
      \item The experimental observation suggests that the particle removal efficiency increases by increasing the applied potential.

  Overall, this study proposed a successful approach of particle cleaning by impinging of charging the drop on the surface.
     
\end{itemize}

\bibliography{pof}

\end{document}